# On Energy Efficiency of Networks for Composable Datacentre Infrastructures

Opeyemi O. Ajibola, Taisir E. H. El-Gorashi, and Jaafar M. H. Elmirghani

*School of Electronic and Electrical Engineering, University of Leeds, LS2 9JT, United Kingdom*

**ABSTRACT**
This paper evaluates the optimal scale of datacentre (DC) resource disaggregation for composable DC infrastructures and investigates the impact of present day silicon photonics technologies on the energy efficiency of different composable DC infrastructures. We formulated a mixed integer linear programming (MILP) model to this end. Our results show that present day silicon photonics technologies enable better network energy efficiency for rack-scale composable DCs compared to pod-scale composable DCs despite reported similarities in CPU and memory resource power consumption.

## 1. INTRODUCTION

Datacentres (DCs) are at the heart of present day digital transformation, because they provide platforms that support digital transformation enabling workloads. These workloads are heterogeneous and are designed for social, enterprise, big data analytics and mobile applications which require unique composition of computation resources for peak performance. This leads to the heterogeneity of nodes in traditional DC and classification of DCs into private enterprise, university and cloud DCs [1]. In recent times, there has been an increase in the migration of applications to DCs to leverage on the pay as you go computing model made possible by cloud computing. Likewise, new workloads are constantly emerging because of the wide uptake of digital transformation across multiple sectors of the global society. To compensate for the growth in the number applications and to support the required quality of service (service level agreements), proportionate increase in size, number and geographical distribution of DCs is required [2].

The architecture adopted in traditional DCs which is based on the interconnection of monolithic servers (with integrated compute, memory and network resource) is plagued with several challenges such as fragmented resource utilization leading to cost and power inefficiencies[3], [4], poor upgrade efficiency leading to high total cost of ownership, integration of node and resource proportionalities, and node-limited resource utilization boundaries that limit scalability [5]. The rigid architecture and utilization boundaries of hardware resources in traditional DCs also limits the gains attainable via the adoption of compute virtualization. Hence, a novel approach is needed in the design of next generation DCs for all categories of applications to ensure business and environment sustainabilities at the same time [5]. These next generation DC infrastructures are required to accommodate present and future growth in the migration of heterogeneous workloads onto on premise and public DCs and the emergence of disruptive workloads for IoT and big data applications. A composable DC infrastructure enabled by high throughput and ultra-low latency optical communication network promises to deliver these requirements. Prior to this work, extensive studies aimed at improving energy efficiency of core network infrastructures connecting geographically distributed DCs have been performed by authors in [6]–[16]. However, this paper focuses on the energy efficiency of intra-DC networks.

In this paper, we investigate and compare the performance of composition of DC infrastructure at rack-scale and pod-scale to deduce the optimal scale of DC infrastructure composition. We also study the effect of composition of DC infrastructure at rack-scale and pod-scale on throughput and energy efficiency of present day networking technologies. Finally, we evaluate the feasible gains in energy efficiency of network infrastructure via the adoption of next generation of silicon photonics technologies in composable DCs. These evaluations, studies and investigations are performed via the formulation of a mixed integer linear programming (MILP) model to represent composable DC infrastructure at different scales.

## 2. COMPOSABLE DATACENTRE INFRASTRUCTURES

The concept of composable DC infrastructure leverages on the growth in the throughput of communication networking technologies and software definition of DC resources. In the composable DC infrastructure, traditional DC servers are disaggregated into pools of homogeneous resources i.e. compute/memory/storage resources. These disaggregated resources are composed, decomposed and recomposed over a high throughput and low-latency network infrastructure on-demand using a centralized software defined orchestration and management layer. With this ability, a dynamic infrastructure with fluid pool of resources that can be composed on-demand to support the resource requirements of a unique workload of any type and size for a defined duration is enabled. This type of infrastructure addresses challenges of traditional DC architecture by improving resource modularity and lifecycle management, removing the bottleneck of workload-specific (bespoke) hardware infrastructure in DCs, agile and rapid composition of DC resources for emerging applications, efficient utilization of DC resources and improved energy and cost efficiencies.

The feasibility of composable infrastructures relies on the ability to physically separate DC resources, the ability to aggregate these resources to form a temporal system and the presence of a communication network with sufficient bandwidth and appropriate latency. Our work here focuses on the first and third of these requirements. Disaggregation of traditional DC resource can be performed at different scales i.e. rack-scale [5], [17] and pod-scale [18], [19] as shown in Figure 1(a) and 1(b) respectively. Likewise, the network interconnecting disaggregated resources can be electrical, hybrid (electro-optical) or all optical. Therefore, a careful evaluation of these options is required in the design of composable infrastructure. Such evaluation must also consider energy efficiency in addition to other metrics (resource utilization and quantity of resource) to ensure that environmental sustainability is also considered in the design process.

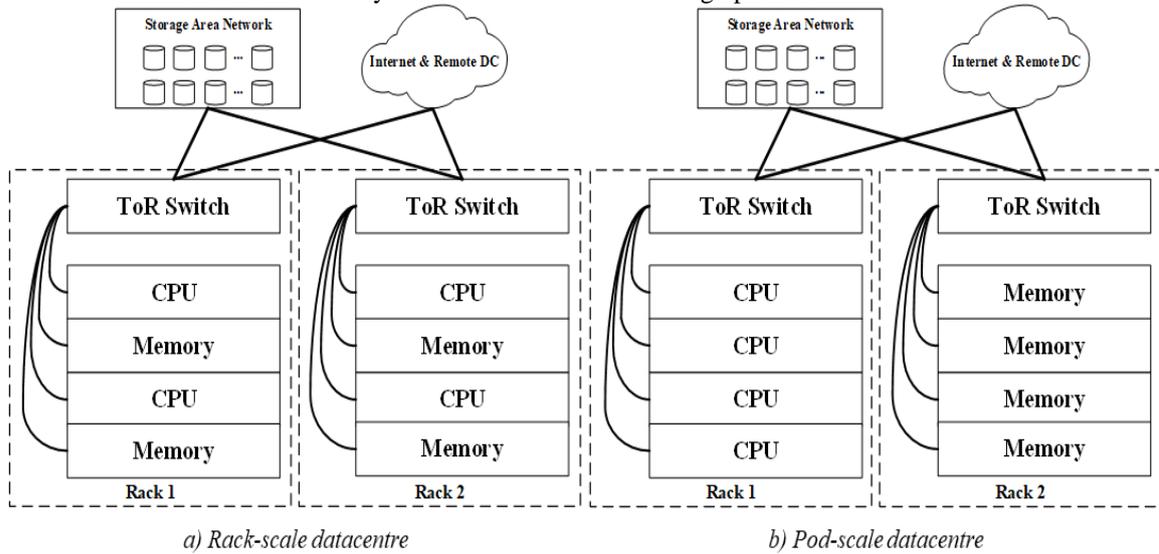

*Figure 1. Scales of DC Resource Disaggregation*

We formulate a MILP model that enables the composition of DC resources into rack-scale and pod-scale DCs as shown in Figure 1. As in Figure 1(a), at rack-scale, a rack in the DCs is a pool which has many CPUs, and many memory modules. At pod-scale (Figure 1(b)), a rack in the DC is a pool; however, in this case, each rack only has many CPU or many memory modules. For completeness, we also model a traditional DC server (TS) as pool comprising of single unit of CPU and memory modules. Under all DC architectures, IO resources are integrated into the DC's communication fabric and we do not adopt any specific communication network architectures for simplicity. The MILP model optimally places workloads with CPU and memory resources demand and inter-resource network traffic and north-south traffic while minimizing the total CPU, memory and network power consumption as given in Eq. (1).

$$Minimize: Total\ CPU\ PC + Total\ Memory\ PC + Total\ Network\ PC \tag{1}$$

The power consumption of CPU and memory resources are estimated by considering the idle power consumption and the load proportionate power consumption over the resources dynamic power range as defined in [18]. The network traffic in each tier (intra-rack or inter-rack or inter-DC tiers) of the DC communication fabric is calculated based on the knowledge of workloads' resource demands placement across the DC. The sum of traffic in each tier is multiplied by the corresponding energy per bit value of the tier (adopted from [20] given in Table 1) to calculate the corresponding network power consumption in that tier of the DC fabric.

*Table 1. Model input parameters.*

| Parameters | Value |
| --- | --- |
| CPU capacity, peak power and dynamic power range | 3.6 GHz, 130 W and 30% |
| Memory capacity, peak power and dynamic power range | 24 GB, 40 W and 30% |
| On-board network fabric energy per bit | 1 pJ/bit |
| Inter-rack network fabric energy per bit | 35 pJ/bit |
| Rack backplane network fabric energy per bit | 25 pJ/bit |
| Inter-DC network fabric energy per bit | 500 pJ/bit |

## 3. COMPARISION OF COMPOSABLE DATACENTRE INFRASTRUCTURE

We consider a small DC with 20 CPU and 20 memory modules of 3.6 GHz and 24 GB capacities respectively to compare the composable data centre architectures. We generated 20 monolithic workloads within the workload resource demand intensity range defined in Table 2 using uniform distribution while CPU-memory, CPU-IO and Memory-IO traffic are fixed to ensure fair comparison between DC architectures. The results from the model can aid the evaluation of CPU and memory resources power consumption, number of active DC resources and average active resource utilization, but our focus here is the evaluation of network traffic in the different tiers of the DC network and the resulting power consumption of an uncapacitated network fabric. CPU and memory resources are composed into traditional DC architecture, disaggregated rack-scale DC and disaggregated pod-scale DC. Note that the term "IOinPool" implies that the total input/output (IO) traffic is restricted to intra-server or intra-rack (i.e. intra-pool) network fabric, "IOinPod" implies that the total IO traffic is restricted to inter-server or inter-rack (inter-pod) network fabric while "PodIO" implies that the total IO traffic is leaving and entering a pod to and from the internet.

*Table 2. Workloads resource demand intensity.*

| Resource Demand \ Workload type | Resource intensity |
|---|---|
| CPU Demand (GHz) | 1-3 |
| Memory Demand (GB) | 5-8 |
| Uplink/Downlink CPU-Memory traffic (Gbps) | 120/100 |
| Uplink/Downlink CPU-IO traffic (Gbps) | 2/1 |
| Uplink/Downlink Memory-IO traffic (Gbps) | 2/1 |

Note that the number of active CPU modules and CPU power consumption are exceptionally high under all DC architecture because the input workloads are CPU intensive. On the other hand, if memory intensive workloads are adopted, more memory modules will be required. The high intensity of CPU resource demand also restricts the ability to consolidate CPU resource demands under all DC architectures. As a result, the same total CPU power consumption (TCPC) values are reported under all DC architectures as shown in Figure 2(a). Hence, workloads' resource intensity is a major factor that influences the resource power consumption and number of active CPU and memory modules in disaggregated DCs. Close observation of the total memory power consumption (TMPC) under different DC architectures reveals the benefits of disaggregating DC resources. Because the range of memory resource demand intensity given in Table 2 is not intensive relative to the defined memory module capacity, significant (49%) savings in total memory power consumption is achieved over traditional DC architecture via the adoption of rack-scale or pod-scale composable DCs as shown in Figure 2(a). This is enabled by increased consolidation memory resource demands onto fewer active memory modules as shown in Figure 2(b). These results show that disaggregation at rack-scale delivers the same CPU and memory power consumption as disaggregation at pod-scale using equal number of CPU and memory modules as shown in Figure 2.

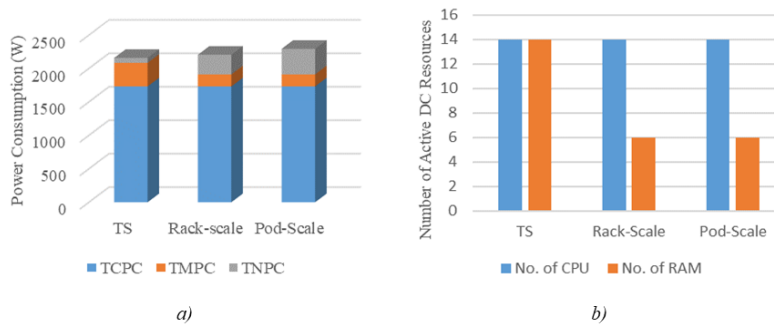

*a) b)*

*Figure 2: (a) Total DC power consumption and (b) Number of active DC resources.*

However, rack-scale based DC delivers better performance in terms of total network power consumption (TNPC) than disaggregation at pod-scale using conventional networking technologies. The total network power consumption of pod-scale DC is 30% higher than that of a rack-scale DC using present day silicon photonic technologies. This is because bandwidth intensive memory access is restricted to on-board and rack backplane fabrics (where network energy efficiency is higher) in rack-scale DC. On the other hand, memory access traffic also traverses the inter-rack fabric in pod-scale DC as shown in Figure 3(a). The results also highlight the pros and cons of traditional DC server architecture relative to CPU, memory and network resources power consumption. For example, 300% increase in total network fabric power consumption is recorded because of disaggregating traditional DC servers at rack-scale as shown in Figure 2(a). The resulting network power consumption at different tiers of the network fabric further illustrates the effect of disaggregation on the overall network power consumption as shown in Figure 3(b). While network power consumption resulting from total IO traffic leaving and entering the pod is similar under all DC architectures, the network power consumption as a result of IO traffic within intra-pool and intra-pod fabric varies with the choice of DC architecture. In the rack-scale DC, bandwidth intensive memory access IO traffic is restricted to the intra-rack backplane for each pool (rack). This aids reduction of total network power consumption relative to the pod-scale DC architecture where memory access IO traffic is present in the inter-rack fabric with lower energy efficiency as illustrated in Table 1. The difference in the energy efficiency of rack back-plane and inter-rack fabrics is responsible for the reported 30% increase in total network power consumption using conventional silicon photonics technologies. Improvements in next-generation silicon photonics technologies that address the energy efficiency gap between these tiers of DC network fabric can reduce the inequality in power consumption of rack-scale and pod-scale DCs.

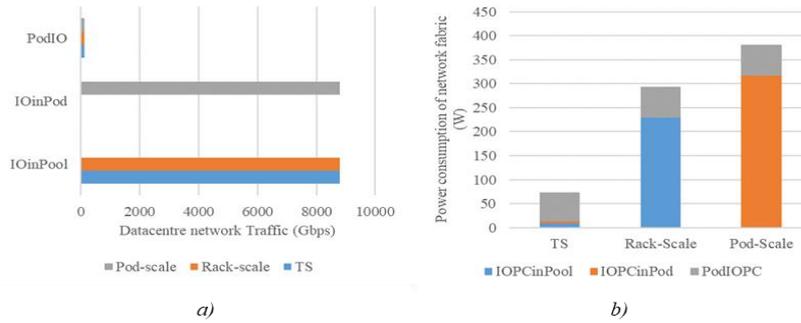

*Figure 3: (a) Traffic in DC network fabric and (b) Power consumption of DC network fabric using silicon photonics.*

## 4. CONCLUSIONS

In this paper, we have formulated a MILP model to evaluate the performance of two parallel scales of resource disaggregation in a composable DC infrastructure. Our results showed that rack-scale disaggregation delivers optimal CPU and memory resource utilization and energy efficiencies while pod-scale disaggregation delivers similar performance at additional cost in the network power consumption using conventional silicon photonic technologies. Advances in silicon photonic technologies are required to address the gap between the energy efficiency of different tiers of network fabric in composable DCs in order to elevate the fears associated with increased network power consumption resulting from disaggregation beyond rack-scale. However, the impact of disaggregation beyond rack-scale on other performance requirements such as resource access latency are also worth exploring.

## ACKNOWLEDGEMENTS

The authors would like to acknowledge funding from the Engineering and Physical Sciences Research Council (EPSRC), INTERNET (EP/H040536/1) and STAR (EP/K016873/1) projects. The first author would like to acknowledge the support of the Petroleum Technology Trust Fund (PTDF), Nigeria, for the Scholarship awarded to fund his PhD. All data are provided in full in the results section of this paper.